\documentclass[aps,pre,onecolumn,showpacs]{revtex4}
\usepackage{amsmath}
\usepackage{epsfig}
\usepackage{amssymb}
\usepackage{dcolumn}
\usepackage{graphicx}
\usepackage{dcolumn}
\newcommand{\sech}{\mathop{\mbox{sech}}}
\newcommand{\sn}{\mathop{\mbox{sn}}}
\newcommand{\dn}{\mathop{\mbox{dn}}}
\newcommand{\cn}{\mathop{\mbox{cn}}}

\newcommand{\tg}{\tilde g}

\usepackage[usenames]{color}

\begin{document}
\date{\today}

\title{Solitary waves in coupled nonlinear Schr\"odinger equations with spatially inhomogeneous nonlinearities}

\author{Juan Belmonte-Beitia}
\email{juan.belmonte@uclm.es}

\affiliation{Departamento de Matem\'aticas, Escuela T\'ecnica
Superior de Ingenieros Industriales, and Instituto de Matem\'atica Aplicada a la Ciencia y la Ingenier\'{\i}a (IMACI),
Universidad de Castilla-La Mancha, 13071 Ciudad Real, Spain.}

\author{Valeriy Brazhnyi}
\email{brazhnyy@gmail.com}

\affiliation{Centro de F\'{\i}sica do Porto, Faculdade de Ci\^encias, Universidade do Porto, R. Campo Alegre 687, Porto 4169-007, Portugal.}

\author{V\'{\i}ctor  M. P\'erez-Garc\'{\i}a}
\email{victor.perezgarcia@uclm.es}
\homepage{http://matematicas.uclm.es/nlwaves}

\affiliation{Departamento de Matem\'aticas, Escuela T\'ecnica
Superior de Ingenieros Industriales, and Instituto de Matem\'atica Aplicada a la Ciencia y la Ingenier\'{\i}a (IMACI),
Universidad de Castilla-La Mancha, 13071 Ciudad Real, Spain.}

\begin{abstract}
Using Lie group theory we construct explicit solitary wave solutions of coupled nonlinear Schr\"odinger systems with  spatially inhomogeneous nonlinearities. We present the general theory,  use it to construct different families of explicit solutions and study their linear and dynamical stability.
\end{abstract}


\maketitle

 \section{Introduction}  
 
 The nonlinear Schr\"odinger equation (NLS) is one of the universal models of mathematical physics arising in many different physical scenarios \cite{Sulem,Vazquez}.
 In recent years there has been an enormous interest on the study of NLS equations with nonlinear coefficients depending either on space, time or both. The motivation comes from the applications of the model to the fields of  Bose-Einstein condensates (BECs) and nonlinear optics. 
 In the first field of application, the so-called Feschbach resonance management allows for a precise control of the atomic interactions responsible for the strength of nonlinearities. This has been experimentally implemented leading to time dependent nonlinear coefficients and would allow for the generation of space or space-time dependent nonlinearities. 
 
Many different types of nonlinear waves have been studied theoretically, experimentally or numerically in the context of BECs (see e.g. the reviews \cite{Panos,VN}). However the consideration of inhomogeneous nonlinearities has led to the prediction of many remarkable nonlinear phenomena in  the last few years, either for 
time dependent   \cite{T1,T2,T3,T4,T5,T6,T7,T8,T9,T10,T11} or space dependent nonlinear coefficients 
 \cite{V1,V2,V3,V4,V41,V5,V6,V7,V9,chinos,V10,V11,V11a,V13,V14,V15,V16,V17,V18,V21,V20,V22}.
 
 Although the model equations of interest are not integrable in general,  there have been efforts to construct families of exact solutions of the NLS with spatially inhomogeneous coefficients. Specifically, group-theoretical methods based on Lie symmetries have been used to obtain solutions of NLS type equations in Refs. \cite{V9,chinos,V15,V20,O1,O2,O3}. However,  up to now the applications  of the technique have been restricted to scalar systems. The purpose of this paper is to extend the method and construct exact solutions for multicomponent  systems. These systems  are of an enormous physical interest because of the extra degrees of freedom provided by the existence of two independent subsystems, e.g. in term of application to multicomponent BEC the model describes two hyperfine levels of an atomic BEC. Exact soliton solutions for such systems in the case of homogeneous in time and space nonlinearities have been reported in many papers (see e.g. \cite{sol0,sol1,sol2,sol4,sol5,sol6,sol7} and references therein). 
Very recently in \cite{cheng} the dynamics of the multicomponent BEC with spatially modulated nonlinearity was treated by means of the the variational approach.
Here we will  go beyond  previous studies and use Lie symmetry method to construct exact bright-bright, dark-dark and dark-bright solutions of the two-component system with spatially inhomogeneous nonlinearities and investigate their linear and dynamics stability.
  
The paper is organized as follows. First, in Section \ref{Sec2}, we introduce the general theory of the Lie symmetry analysis for our model problem: a system of two coupled nonlinear Schr\"odinger  equations with spatially inhomogeneous nonlinearities. In Section \ref{Sec3}, we use the method  to construct explicit solutions for systems without external potentials $(V_{j}(x)=0$, $j=1,2$) and study the linear and dynamical stability of the obtained solutions. In Section \ref{Sec4}, we present some families of exact solutions for inhomogeneous coupled nonlinear Schr\"odinger equations with external potentials quadratic in space and propose their stability analysis. Next, in Section \ref{Sec5}, we discuss the construction of dark-bright soliton solutions and study their stability. Finally, in Section \ref{Sec6}, we summarize our conclusions. The theory of the linear stability analysis for the system is presented in Appendix \ref{ApenA}.
 
 \section{General theory}
 \label{Sec2}
 
 \subsection{Model to be studied}
 
 In this paper we will construct and study the stability of solutions of the following  coupled spatially inhomogeneous nonlinear Schr\"odinger equations (CINLS)
  \begin{subequations}
   \label{GP}
  \begin{eqnarray}
i\frac{\partial \psi_1}{\partial t} &=&-\frac{\partial^{2}\psi_{1}}{\partial x^{2}}+V_{1}(x)\psi_{1}+\left(g_{11}(x)|\psi_{1}|^{2}+g_{12}(x)|\psi_{2}|^{2}\right)\psi_{1},\\
  i\frac{\partial \psi_2}{\partial t}&=&-\frac{\partial^{2}\psi_{2}}{\partial x^{2}}+V_{2}(x)\psi_{2}+\left(g_{21}(x)|\psi_{1}|^{2}+g_{22}(x)|\psi_{2}|^{2}\right)\psi_{2},
\end{eqnarray}
\end{subequations}
 for known real potential functions $V_{j}(x)$, and spatial modulations of the nonlinearities described by the real functions $g_{jk}(x)$, $j,k=1,2$. 
 In applications to BEC, $\psi_1$ and $\psi_2$ are complex wave functions describing two hyperfine levels (the extension to different atomic species is simple, corresponding to the addition of different masses)  and describe the dynamics of coupled quasi-one dimensional BECs in the mean field limit  with spatially inhomogeneous interactions.
 
In the following we will be interested in solitary wave solutions of Eqs. (\ref{GP}) of the form $\psi_j(x,t)= u_j(x) e^{-i\lambda_j t}$, where $u_j(x)$ are real functions ($j=1,2$). Upon substitution in Eqs. \eqref{GP} we get
system of stationary equations
   \begin{subequations}
    \label{estacionario}
  \begin{eqnarray}
 \lambda_{1} u_{1}&=&-\frac{d^{2}u_{1}}{dx^{2}}+V_{1}(x)u_{1}+\left(g_{11}(x)u_{1}^{2}+g_{12}(x)u_{2}^{2}\right)u_{1},\\
  \lambda_{2} u_{2}&=&-\frac{d^{2}u_{2}}{dx^{2}}+V_{2}(x)u_{2}+\left(g_{21}(x)u_{1}^{2}+g_{22}(x)u_{2}^{2}\right)u_{2}.
\end{eqnarray}
\end{subequations}
The construction of families of exact solutions of Eqs. (\ref{estacionario}) is the main goal of this paper.
 
 \subsection{Lie symmetry method for Eqs. (\ref{estacionario})}
 
 In this section we will extend the method of Ref. \cite{V9} to the CINLS given by Eqs. (\ref{estacionario}). The methodology is similar although the calculations are more complicated, but we will show that it is still possible  to extend the method for this more complicated situation.
 
 It is said that  a second-order differential system
\begin{equation}\label{ODE}
N_{j}(x,u_j,u_{j,x},u_{j,xx})=0,
\end{equation}
possesses a Lie point symmetry  \cite{Bluman} of the form
\begin{equation}
M=\xi(x,u_{1},u_{2}) \frac{\partial}{\partial x}+\eta_1(x,u_{1},u_{2})\frac{\partial}{\partial u_1}+\eta_2(x,u_1,u_2)\frac{\partial}{\partial u_2},
\end{equation}
 if the action of the second extension of $M$, i.e.  $M^{(2)}$ on $N_{j}$, gives zero whenever (\ref{ODE}) is satisfied, i.e.
\begin{multline}
M^{(2)} N_{j}(x,u_{j},u_{j,x},u_{j,xx}) = 
\left[\xi(x,u_{1},u_{2})\frac{\partial}{\partial x}+
\eta_1(x,u_{1},u_{2})\frac{\partial}{\partial u_{1}} + \right. \\
\left. \eta_2(x,u_{1},u_{2})\frac{\partial}{\partial u_{2}}+  \eta_{1}^{(1)}(x,u_{1},u_{2},u_{1x})\frac{\partial}{\partial u_{1,x}}+ \eta_{2}^{(1)}(x,u_{1},u_{2},u_{2,x})\frac{\partial}{\partial u_{2,x}}\right.  \\ \left.
+\eta_{1}^{(2)}(x,u_{1},u_{2},u_{1,xx})\frac{\partial}{\partial u_{1,xx}} + 
 \eta_{2}^{(2)}(x,u_{1},u_{2},u_{2,xx})\frac{\partial}{\partial u_{2,xx}}\right] N_{j}(x,u_{j},u_{j,x},u_{j,xx}) = 0  %\right. \\ \left.
\end{multline}
where $u_{j,x}\equiv du_{j}/dx$ and $u_{j,xx}\equiv d^{2}u_{j}/dx^{2}$, with $j=1,2$.

In our problem, $N_{j}(x,u_{j},u_{j,xx})$, $j=1,2$ are given by
\begin{subequations}
\begin{eqnarray}
N_{1}(x,u_{1},u_{2},u_{1,xx})&=&u_{1,xx}-f_{1}(x,u_{1},u_{2}),\\
N_{2}(x,u_{1},u_{2},u_{2,xx})&=&u_{2,xx}-f_{2}(x,u_{1},u_{2}),
\end{eqnarray}
\end{subequations}
where 
\begin{subequations}
\begin{eqnarray}
f_{1}(x,u_{1},u_{2})&=&\left[V_{1}(x)-\lambda_1+g_{11}(x)u_{1}^{2}+g_{12}(x)u_{2}^{2}\right] u_{1},\\
f_{2}(x,u_{1},u_{2})&=&\left[V_{2}(x)-\lambda_2+g_{21}(x)u_{1}^{2}+g_{22}(x)u_{2}^{2}\right] u_{2},
\end{eqnarray}
\end{subequations}
and the action of the operator $M^{(2)}$ on $N_{1}$ leads to 
\begin{subequations}
\begin{eqnarray}
\label{relaciones-q}
&&\xi_{u_{1}u_{1}}=0,
\\
&&\xi_{u_{1}u_{2}}=0,
\\
&&\xi_{u_{2}u_{2}}=0,
\\
&&\eta_{1,u_{2}u_{2}}=0,
\\
&&\eta_{1,u_{1}u_{2}}=\xi_{u_{2,x}},
\\
&&\eta_{1,u_{2}x}=\xi_{u_2}f_{1},
\\
&&\eta_{1,u_{1}u_{1}}=2\xi_{xu_{1}},
\\
&&2\eta_{1, xu_{1}}-\xi_{xx}-3\xi_{u_1}f_{1}-f_{2}\xi_{u_2}=0,
\\
&&\eta_{1,xx}+\left(\eta_{1,u_{1}}-2\xi_{x}\right)f_{1}-\xi f_{1,x}-\eta_1 f_{1,u_1}-\eta_ 2 f_{1,u_2}+f_2\eta_{1,u_2}=0.
\end{eqnarray}
\end{subequations}

In a similar way, we get equations for the action of $M^{(2)}$ on $N_{2}$ (we omit the details). Solving the previous equations, we find that the only Lie 
point symmetries of Eq. (\ref{estacionario}) are of the form
\begin{equation}
\label{simetria}
M=c(x)\frac{\partial}{\partial x}+m(x)u_{1}\frac{\partial}{\partial u_{1}}+m(x)u_{2}\frac{\partial}{\partial u_{2}},
\end{equation}
where
\begin{subequations}
\label{relaciones}
\begin{eqnarray}
&&g_{jk}(x)  =  \tg_{jk}/c(x)^{3}, \label{nonlincoef}\\
&&m(x)= c'(x)/2, \\\label{relacionesc}
&&-c(x)V'_{1}(x)+\frac{c'''(x)}{2}-2c'(x)\left(V_{1}(x)-\lambda_1\right)=0, \\\label{relacionesd}
&&V_{2}(x)=V_{1}(x)+ K/c^{2}(x)+\lambda_{2}-\lambda_{1}. 
\end{eqnarray}
\end{subequations}
Here $\tg_{jk}=$constants and $K$ an arbitrary real number. 

 It is known \cite{Leach}, that the invariance of the energy is associated to the translational invariance. 
  The generator of such a transformation is of the form $M=\partial/\partial x$. To use this fact, we define the transformation
\begin{equation}
\label{transformaciones}
X=h(x),\qquad U_{1}=v(x)u_{1},\qquad U_{2}=v(x)u_{2},
\end{equation}
where $h(x)$ and $v(x)$ will be determined by requiring 
 that a conservation law of energy type $M=\partial/\partial X$ exists 
in the canonical variables. Using Eqs. (\ref{simetria}) and  \eqref{transformaciones}, we get
\begin{equation}
h(x)  =  \int_{0}^{x} \frac{1}{c(s)}ds,\qquad  v(x)  =  c(x)^{-1/2}.
\end{equation}
 Rewriting Eqs. (\ref{estacionario}) in terms of
$U_{j}=c^{-1/2}(x)u_{j}$, $j=1,2$ and 
$X=\int_{0}^{x}1/c(s) ds$ we get
\begin{subequations}
\label{homogenea1}
\begin{eqnarray}
-\frac{d^{2}U_{1}}{dX^{2}}+\left(\tg_{11}U_{1}^{2}+\tg_{12}U_{2}^{2}\right)U_{1}&=&E_{1}U_{1},\\
-\frac{d^{2}U_{2}}{dX^{2}}+\left(\tg_{21}U_{1}^{2}+\tg_{22}U_{2}^{2}\right)U_{2}&=&E_{2}U_{2},
\end{eqnarray}
\end{subequations}
where 
\begin{subequations}
\label{E1}
\begin{eqnarray}
E_{1}&=& \left(\lambda_{1}-V_{1}(x)\right) c(x)^{2}-\tfrac{1}{4}c'(x)^{2}+ \tfrac{1}{2}c(x)c''(x),\\
E_{2}&=& \left(\lambda_{2}-V_{2}(x)\right) c(x)^{2}-\tfrac{1}{4}c'(x)^{2}+ \tfrac{1}{2}c(x)c''(x),
\label{E2}
\end{eqnarray}
\end{subequations}
 are constants. Thus, in the new variables we obtain nothing else but system of NLS equations \emph{without external potential and with spatially homogeneous nonlinearities}.  Of course not any choice of $V_{j}(x)$ and $g_{jk}(x)$, $j,k=1,2$, leads to the existence of a Lie symmetry or an appropriate canonical transformation since the function $c(x)$ must be sign definite for $U_{j}$,  $j=1,2$ and $X$ to be properly defined. 
In the specific case when all nonlinear coefficients are equal, i.e. $\tg_{jk}=\tg$ for all $j,k=1,2$, Eqs.  (\ref{homogenea1}) correspond to the time independent Manakov system \cite{sol0}, that has been studied in detail (see, for instance \cite{sol1,sol2,sol4}). 
 
 We can thus use the known solutions of Eqs. (\ref{E2}) to construct solutions of Eqs. (\ref{estacionario}) in the same spirit as in the scalar case (see e.g. Ref. \cite{V6}). Moreover, when both $E_1=E_2=E$ and the restriction
\begin{eqnarray}
\label{cond}
(\tg_{11}-\tg_{21})(\tg_{22}-\tg_{12})>0,
\end{eqnarray}
are satisfied,  Eqs. (\ref{homogenea1}) reduce to the scalar NLS equation
\begin{eqnarray}
\label{homogenea2}
-\frac{d^{2}\Phi}{dX^{2}}+\sigma \Phi^{3}=E \Phi,
\end{eqnarray}
where we have introduced the new field $\Phi$,  related with the original fields through the relation
\begin{equation}
\label{PhitoU}	
U_j=\theta_j \Phi.
\end{equation}
Here $\theta_1=\sqrt{|\tg_{22}-\tg_{12}|/Q}$, and  
$\theta_2=\sqrt{|\tg_{11}-\tg_{21}|/Q }$, being $Q = |\tg_{11}\tg_{22}-\tg_{12}\tg_{21}|$ 
and $\sigma={\rm sgn}[(\tg_{11}\tg_{22}-\tg_{12}\tg_{21})(\tg_{jj}-\tg_{12})]$  which due to the condition (\ref{cond})  does not depend on $j$.
 In that situation we can use the many known solutions of the scalar equation (\ref{homogenea2}) to construct solutions of the vector problem (\ref{estacionario}). 
 Equation (\ref{homogenea2}) has many known exact solutions. Remarkable examples, to be used later in this paper, are
 \begin{subequations}
\begin{eqnarray}
\label{sol_ex_1}	
\Phi&=& \mu\, \tanh\left(\frac{\mu}{\sqrt{2}} X\right), \quad E=2\mu^2, \quad \sigma=1,\\
\label{sol_ex_2}	
\Phi&=&\sqrt{2} \mu\,  \frac{1}{\cosh(\mu X)}, \quad E=-\mu^2, \quad \sigma=-1, \\
\label{sol_ex_3}	
\Phi&=& \mu k \sqrt{2(1-k^2)} \, \frac{{\rm sn}(\mu X, k)}{{\rm dn}(\mu X, k)}, \quad E=\mu^2(1-2k^2), \quad \sigma=-1,\\
\label{sol_ex_4}	
\Phi&=& \sqrt{2}\mu\, {\rm dn}(\mu X, k), \quad E=\mu^2(k^2-2), \quad \sigma=-1.
\end{eqnarray}
\end{subequations}
   
\section{Systems without external potential ($V_1(x)= V_2(x)= 0$)}
\label{Sec3}

 \subsection{General considerations}
 
As a first application of our ideas let us choose $V_{1}(x)=V_{2}(x)=0$. By taking $K=0$ and $\lambda_{2}=\lambda_{1}\equiv\lambda$ in (\ref{relacionesd}) we find from Eq.~\eqref{relacionesc} the equation $ c'''(x)+4\lambda c'(x)=0$, whose general solutions are given by
\begin{subequations}
\begin{eqnarray}
c(x) & = & C_1 \sin \omega x  + C_2 \cos \omega x  + C_3 \ \ ( \lambda > 0), \label{b1b} \\
c(x) & = & C_1 e^{ \omega x} + C_2 e^{-\omega x} + C_3 \ \ \ (\lambda <0), \label{b1a}
 \end{eqnarray} 
  \end{subequations}
 where $\omega = 2 \sqrt{|\lambda|}$. 
 It should be emphasized that the auxiliary function $c(x)$ is a periodic function  for $\lambda>0$ which corresponds to the continue spectrum of the linear eigenvalue problem $d^2m/dx^2 + 4\lambda m=0$ and for $\lambda <0$ (the region where the linear eigenvalue problem does not have periodic solutions) the function $c(x)$ becomes exponentially localized. 
Therefore, in the following subsections we will distinguish two types of solutions of Eqs. (\ref{GP}): corresponding to either periodic or localized nonlinearities.
 
 \subsection{Dark-dark soliton solutions}

Let us first consider the case $\lambda>0$. Then, taking $C_1=0$, $C_2=\alpha$, $C_3=1$,  Eq.~(\ref{b1b}) leads to a periodic dependence of the coefficient, e.g.  $c(x)=1 + \alpha \cos \omega x$, and according to Eq. (\ref{nonlincoef}) the nonlinear coefficients are given by
 \begin{equation}\label{qper}
 g_{jk}(x) =  \tg_{jk} \left(1 + \alpha \cos \omega x \right)^{-3},\quad 
 j,k=1,2.
  \end{equation}
  Although the form  of nonlinear coefficients seems complicated, it is easy to see that 
for small $\alpha$ this nonlinearity is approximately harmonic in space $
 g_{jk}(x) \simeq \tg_{jk} \left(1 - 3 \alpha \cos \omega x \right)$,  $\alpha \ll 1$, $j,k=1,2$, and thus at least in that limit is of direct physical interest.
 %(see e.g. Fig. \ref{prima}(c)). 
 
We can construct our canonical transformation by using Eqs.~(\ref{transformaciones}) and obtain 
 \begin{equation}\label{cuci}
  \tan \left( \frac{ \omega }{ 2 } \sqrt{1-\alpha^2} \; X(x) \right) = 
  \sqrt{ \frac{1-\alpha}{1+\alpha} } \; \tan \frac{\omega x}{2}.
\end{equation}
Using any solution of Eq. (\ref{homogenea1}) 
with $E=E_{1}=E_{2}=\tfrac{1}{4}\omega^{2}\left(1-\alpha^2\right)>0$, where $\lambda=\omega^{2}/4$,
this transformation provides 
solutions of Eq. (\ref{estacionario}) with $g(x)$ given by Eq. \eqref{qper}. 
For instance,  taking the nonlinear coefficients  to satisfy Eq.  (\ref{cond}) and $\sigma=1$ we can construct dark-dark  soliton solutions of Eq. (\ref{estacionario}) of the form
 \begin{equation}\label{brito}
u_{j}(x)=\theta_{j}\left(1+\alpha\cos(\omega x)\right)^{1/2}\sqrt{2}\mu \tanh\left(\mu X(x)\right),
 \end{equation}
where $\mu=\sqrt{E}=\frac{1}{2}\omega\sqrt{1-\alpha^{2}}$
%
%\begin{equation}
%\mu=\frac{1}{2}\omega\sqrt{\frac{1}{2}(1-\alpha^{2})},\quad a_{1}=\mu\sqrt{2\frac{\tg_{12}-\tg_{22}}{\tg_{12}\tg_{21}-\tg_{11}\tg_{22}}},\quad a_{2}=\mu\sqrt{2\frac{\tg_{21}-\tg_{11}}{\tg_{12}\tg_{21}-\tg_{11}\tg_{22}}}
%\end{equation}
%
and $X(x)$ is defined by Eq. \eqref{cuci}. 
We want to emphasize that this is only a simple example of the many possible solutions 
that can be constructed in such a way.

\subsection{Stability of the dark-dark solution (\ref{brito})}

Let us consider the dynamical stability of the dark-dark soliton solution  (\ref{brito}). 
Here and in all of our calculations to be presented in this paper in order to check dynamical stability we perturb the initial exact solution according to 
\begin{equation}\label{pert}
u_j(x)=u_{j0}[1+\epsilon\cos(q x\pm 1)],
\end{equation}
 where ``$+$'' and ``$-$'' correspond to the first and second components, respectively, $\epsilon$ will be taken as a small number and $q$ is the perturbation wavenumber.

\begin{figure}
 \begin{center}
\epsfig{file=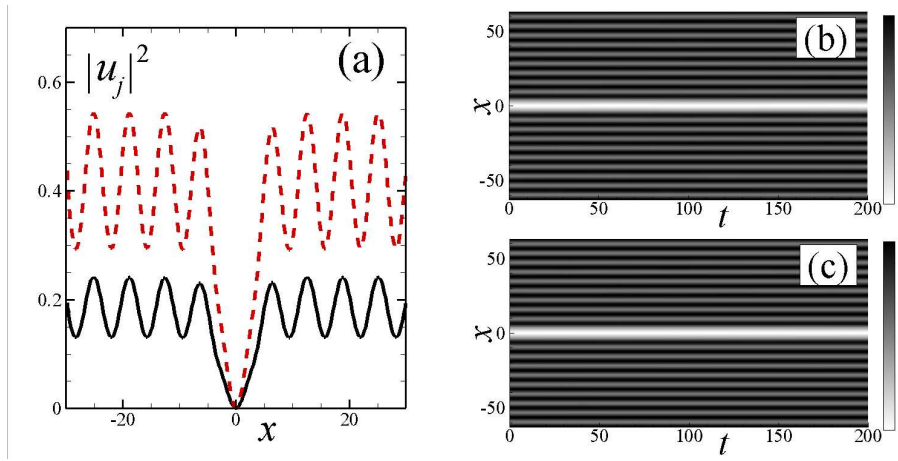,width=12 cm}
\end{center}
 \caption{Dynamics of the dark-dark solution (\ref{cuci}) under a perturbation of the form (\ref{pert}). Shown are (a) the initial density profiles of the first (black solid) and second (red dashed) component. (b,c) Density plot of the time evolution of (b) $ |u_1(x,t)|^2$ (c) $ |u_2(x,t)|^2$.
Parameter values are $g=0.1, g_{11}=1$, $g_{22}=0.5$, $\alpha=0.3$, $\omega=1$, $\epsilon=0.001$, $q=0.01$. \label{dyn_DD1}}
 \end{figure}

\begin{figure}
  \begin{center}
\epsfig{file=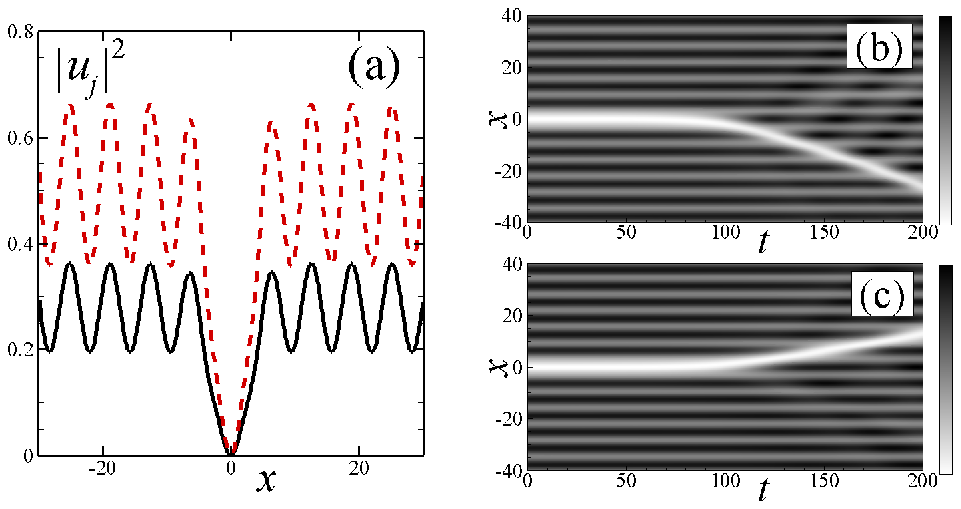,width=12 cm}
\end{center}
 \caption{  Same as in Fig. \ref{dyn_DD1} but for $g=-0.1$.
% Dynamics of the dark-dark solution (\ref{cuci}) under a perturbation of the form (\ref{pert}). Shownare (a) the initial density profiles of the first (black solid) and second (red dashed) component. (b,c) Density plot of the time evolution of (b) $ |u_1(x,t)|^2$ (c) $ |u_2(x,t)|^2$.
%Parameter values are $g=-0.1, g_{11}=1$, $g_{22}=0.5$, $\alpha=0.3$, $\omega=1$, $\epsilon=0.001$, $q=0.01$. 
\label{dyn_DD2}} 
\end{figure}

For the analysis we will fix the nonlinearity for the first component $g_{11}=1$ and change the nonlinearity for the second component in the interval $0<g_{22}<1$ and the intercomponent interaction $g_{12}=g_{21}=g$ in the interval $-\sqrt{g_{22}}<g<g_{22}$.
Our results show that 
 for small $\alpha\ll 1$ the dark-dark solution is dynamically stable in the full range $0<g<g_{22}$ (see Fig. \ref{dyn_DD1}). However, in the interval $-\sqrt{g_{22}}<g<0$ the dark-dark solutions are unstable (see Fig. \ref{dyn_DD2}).
  
%We have also found that by increasing the parameter $\alpha$  defining the strength (depth) of the periodic nonlinearity it is possible to increase the lifetime of the dark-dark solution.

\begin{figure}
 \begin{center}
\epsfig{file=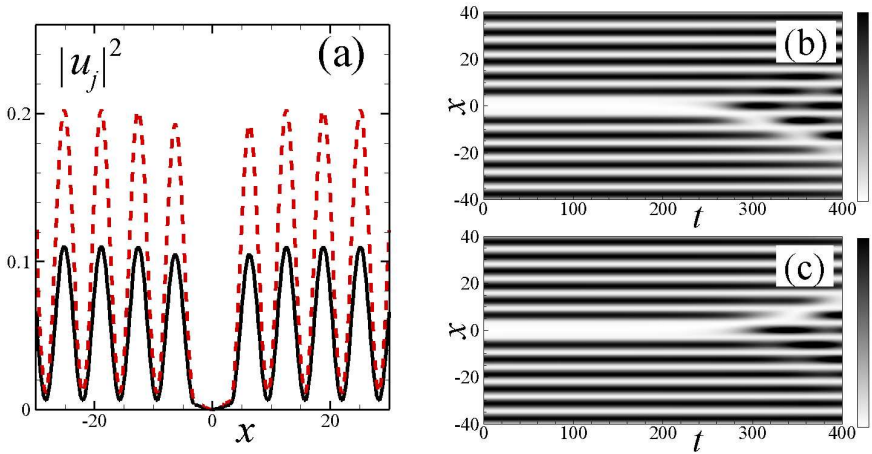,width=12 cm}
\end{center}
 \caption{ Same as in Fig.\ref{dyn_DD1} but for $\alpha=0.9$.
 %Dynamics of the dark-dark solution (\ref{cuci}) under a perturbation of the form (\ref{pert}). Shownare (a) the initial density profiles of the first (black solid) and second (red dashed) component. (b,c) Density plot of the time evolution of (b) $ |u_1(x,t)|^2$ (c) $ |u_2(x,t)|^2$.
%Parameter values are $g=-0.1, g_{11}=1$, $g_{22}=0.5$, $\alpha=0.3$, $\omega=1$, $\epsilon=0.001$, $q=0.01$. 
\label{dyn_DD_alph_09}}
 \end{figure}

Also we have checked that by increasing the amplitude $\alpha$ of the nonlinear inhomogeneity we found the instabilities take longer times to develop. As one can see in Fig. \ref{dyn_DD_alph_09} for $\alpha=0.9$ the instability occurs near $t\approx 250$ while in the previous example for $\alpha=0.3$ shown in Fig. \ref{dyn_DD2} the solution breaks around $t \approx 90$.

\subsection{Bright-bright soliton solutions}
\label{bb}

Let us now consider the case $\lambda <0$ defined by Eq. (\ref{b1a}). 
In order to simplify the analysis we restrict ourselves to two particular choices of the constants $\omega$, $C_{1}$, $C_{2}$ and $C_{3}$. The first one is $\omega =1$, $C_{1}=C_2=1/2$ and $C_3=0$ in Eq. (\ref{b1a}). 
Thus $ c(x) = \cosh x$ and $\lambda=-1/4$, $E=E_{1}=E_{2}=1/4$,
 being  $\cos X(x) = - \tanh x$, and the nonlinear coefficients are
 \begin{equation}
 g_{jk}=\frac{\tg_{jk}}{\cosh^{3}(x)}
 \end{equation}
 Then $0 \le X \le \pi$, and to meet the boundary conditions $u_{j}(\pm\infty)=0$, one has to impose
$U_{j}(0) = U_{j}(\pi) = 0$. This means that the original infinite domain in Eqs. (\ref{estacionario}) is mapped into a bounded domain for Eqs. (\ref{homogenea1}). By choosing the nonlinear coefficients to satisfy Eq. (\ref{cond}) and $\sigma=-1$ we can use the solution (\ref{sol_ex_3}) to construct vector soliton solutions of the form
\begin{equation}
\label{ellip}
  U_{j}(X) =\theta_{j} \; \mu k \sqrt{2(1-k^{2})} \, \frac{ \mathop{\mbox{sn}}(\mu X, k) }{ \mathop{\mbox{dn}}(\mu X, k) },
  \end{equation}
where $ \mu = \frac{1}{2}\sqrt{\frac{1}{ 1 - 2k^{2}}}$.

The functions $U_{1}(X)$ and $U_{2}(X)$ satisfy $U_{1}(0)=U_{2}(0)=0$ and in order to meet $U_{1}(\pi)=U_{2}(\pi)=0$, the condition $\mu\pi = 2nK(k)$ where $K(k)$ is the elliptic integral $ K(k) = 
  \int_{0}^{\pi} 
\left(1 - k^{2} \sin^{2}\varphi\right)^{-1/2} d\varphi$ must hold. 
Thus, to satisfy the boundary conditions, $k$ 
must be chosen to solve the following equation 
\begin{equation}
4n K\left( k \right) 
  \sqrt{ 1 - 2k^{2} }
= \pi\quad \text{for}\quad  n=1, 2, 3, ...
\label{n}
\end{equation}
 It can be shown that for every integer number $n$ this algebraic equation has only a solution $k_n$ which means that 
there are an infinite number of solutions of Eqs. (\ref{estacionario}) of the form given by Eqs. (\ref{ellip}). Moreover, each of those solutions has exactly $n-1$ zeroes. Then, using  \eqref{transformaciones} we find a solution of Eqs. \eqref{estacionario} of the form
\begin{eqnarray}
\label{ellip_u}
  u_{j}(x) =c(x)^{1/2}U_{j}(X(x)).
  \end{eqnarray}
 
\subsection{Stability of the bright-bright solution (\ref{ellip_u})}

\begin{figure}
   \begin{center}
\epsfig{file=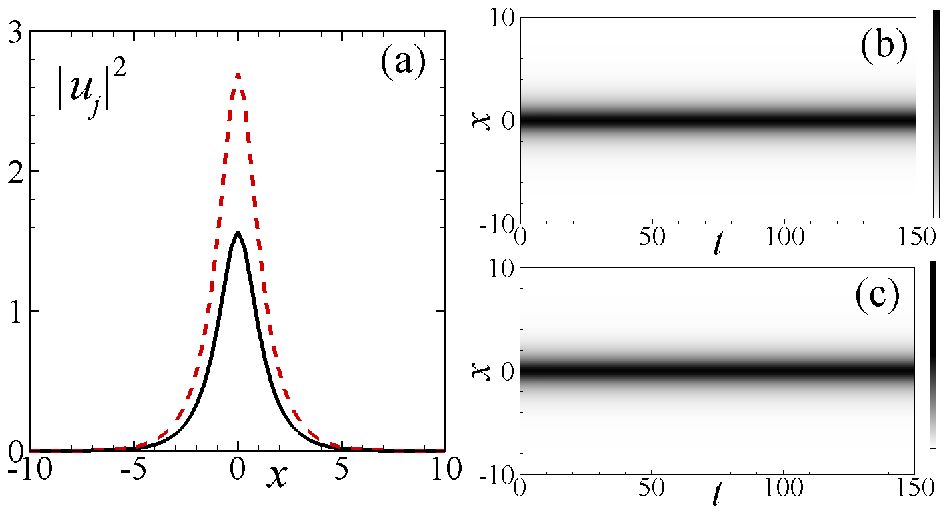,width=12 cm}
  \end{center}
\caption{Dynamics of a bright-bright soliton solution (\ref{ellip_u}) with  $n=1$ under a perturbation of the form (\ref{pert}). Shownare (a) the initial density profiles of the first (black solid) and second (red dashed) component. (b,c) Density plots of the time evolution of (b) $ |u_1(x,t)|^2$ (c) $ |u_2(x,t)|^2$.
Parameter values are $g=0.2$, $g_{11}=-1$, $g_{22}=-0.5$, $\epsilon=0.001$, and $q=0.2$.  \label{dyn_BB1}}
 \end{figure}

\begin{figure}
  \begin{center}
\epsfig{file=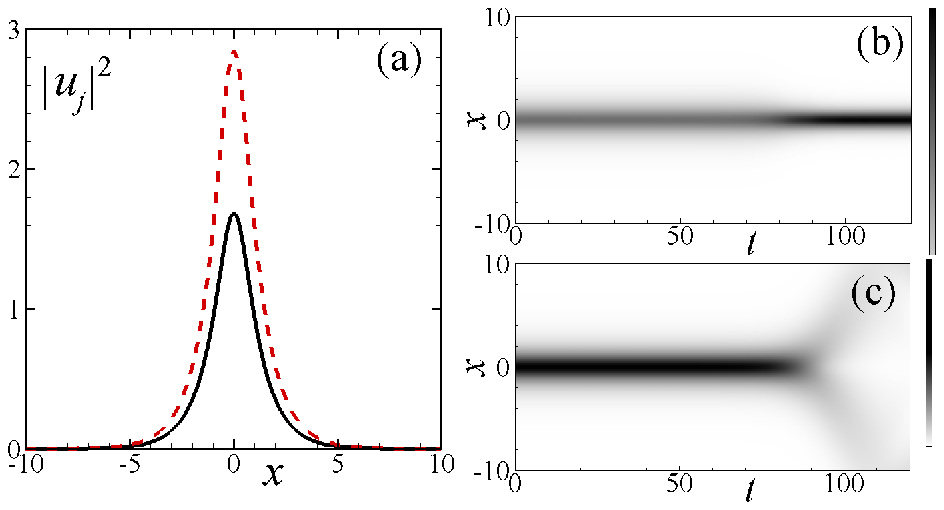,width=12 cm}
  \end{center}
\caption{Same as in Fig.\ref{dyn_BB1} but for $g=0.23$.
\label{dyn_BB2}}
 \end{figure}

Next we investigate the stability of the bright-bright solutions given by Eq. (\ref{ellip_u}) by fixing  $g_{11}=-1$ and choosing different combinations of $(g_{22},g)$ (as in the previous case we consider $g_{12}=g_{21}=g$). 
First we start with the case $n=1$ in Eq. (\ref{n}). 
In this case for each value of $g_{22}$ we find that there exists a  critical value of $g=g_{cr}$ such that for $g<g_{cr}$ the vector solution is stable and for $g>g_{cr}$ the solution becomes  modulationaly unstable. As examples Figs. \ref{dyn_BB1} and  \ref{dyn_BB2} show that the critical value $g_{cr}$ lies between $g=0.2$ and $g=0.23$. 
It should be stressed here that in both regions of existence of the solutions, i.e. $g<g_{cr}$ and $g>g_{cr}$, they  are linearly stable, i.e. there are no negative or complex eigenvalues $\Omega_j$, ($j=1,2$). However dynamically, the solutions for  $g>g_{cr}$ suffer a phase ``separation'' where one of the component extrudes the other component, a phenomenon also observed in related systems Ref. \cite{CBK}.
 
\begin{figure}
   \begin{center}
\epsfig{file=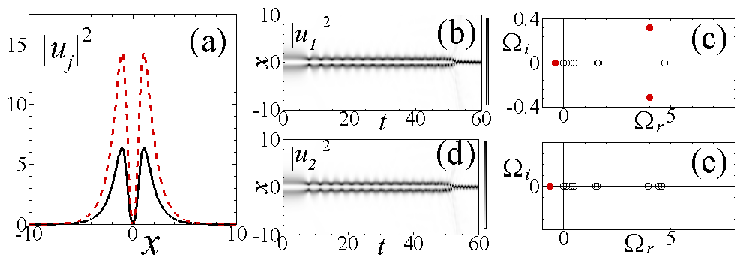,width=12 cm}
   \end{center}
\caption{Dynamics of a $n=2$ bright-bright solution given by Eq.  (\ref{ellip_u}) under a perturbation of the form (\ref{pert}). 
Shown are (a) the initial density profiles of the first (black solid) and second (red dashed) components. 
(b,d) Density plot of the time evolution of (b) $ |u_1(x,t)|^2$ (d) $|u_2(x,t)|^2$.
(c,e) Part of the spectrum of the eigenvalues $\Omega_i(\Omega_r)$ for the first (c) and second (e) components, calculated from Eq. (\ref{eigen_matrix_simpl}). The red dots on the spectrum show the existence of negative and complex eigenvalues.
Parameter values are $g=-0.1$, $g_{11}=-1$, $g_{22}=-0.5$, $\epsilon=0.001$, $q=0.2$.
\label{dyn_BB1_2_g-01}}
 \end{figure}

\begin{figure}
   \begin{center}
\epsfig{file=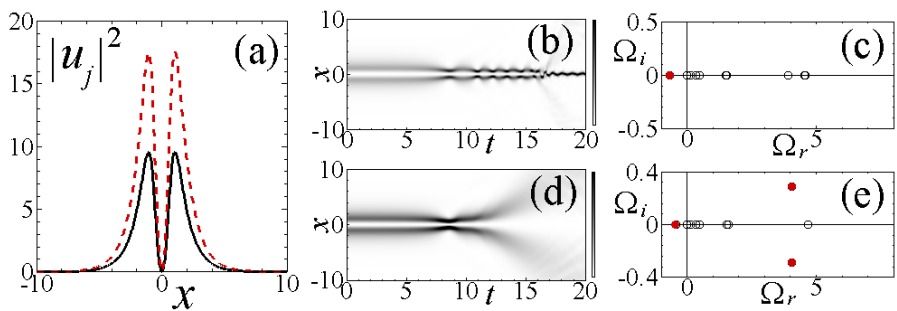,width=12 cm}
   \end{center}
\caption{Same as in Fig. \ref{dyn_BB1_2_g-01} but for $g=0.1$. 
\label{dyn_BB1_2_g_01}}
 \end{figure}

As to solutions with $n=2$ and $3$, the 
 linear stability analysis shows that all high-order solutions are linearly unstable. 
In the case $n=2$, the linearized equations (see Appendix \ref{ApenA}) have one pure negative eigenvalue $\Omega_r={\rm Re}(\Omega)<0$ for both components and also two complex conjugated eigenvalues with non-zero imaginary part $\Omega_i={\rm Im}(\Omega)\ne 0$ for the first component only (see panels (c), (e) in Figs.\ref{dyn_BB1_2_g-01} and \ref{dyn_BB1_2_g_01}).
   
The numerical integration of Eqs. (\ref{GP}) with perturbed initial bright-bright solution with $n=2$ confirms the results of the linear stability analysis. As we can see in panels (b) and (d) of  Figs. \ref{dyn_BB1_2_g-01} and \ref{dyn_BB1_2_g_01}, the dynamics becomes unstable for small times and consists of oscillatory behavior at the initial stages of instability. When the instability develops fully  the initial profile of the solution (having two peaks) degenerates into a one-peak stationary profile corresponding to bright-bright solution with $n=1$ (Fig.\ref{dyn_BB1_2_g-01}). In the case of $g>0$ due to the instability, the second component splits into two wavepackets which move outwards the center while the first component transforms into one-peak bright soliton solution (see Fig.\ref{dyn_BB1_2_g_01}), that is a stable mode of the system.

\begin{figure}
    \begin{center}
\epsfig{file=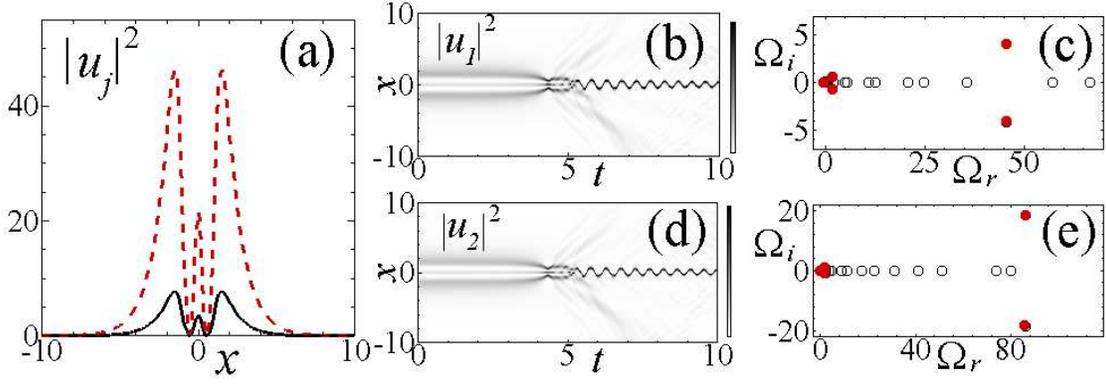,width=15 cm}
   \end{center}
\caption{
Dynamics of a bright-bright solution given by Eq. (\ref{ellip_u}) with $n=3$ under a perturbation of the form (\ref{pert}). 
Shown are (a) the initial density profiles of the first (black solid) and second (red dashed) component. 
(b,d) Density plot of the time evolution of (b) $ |u_1(x,t)|^2$ (d) $|u_2(x,t)|^2$.
(c,e) Part of the spectrum of the eigenvalues $\Omega_i(\Omega_r)$ for the first (c) and second (e) components, calculated from Eq. (\ref{eigen_matrix_simpl}).
 The red dots on the spectrum show the existence of negative and complex eigenvalues.
Parameter values are $g=-0.4$, $g_{11}=-1$, and $g_{22}=-0.5$, $\epsilon=0.001$, $q=0.2$.
 \label{dyn_BB1_3_-04}}
 \end{figure}

\begin{figure}
    \begin{center}
\epsfig{file=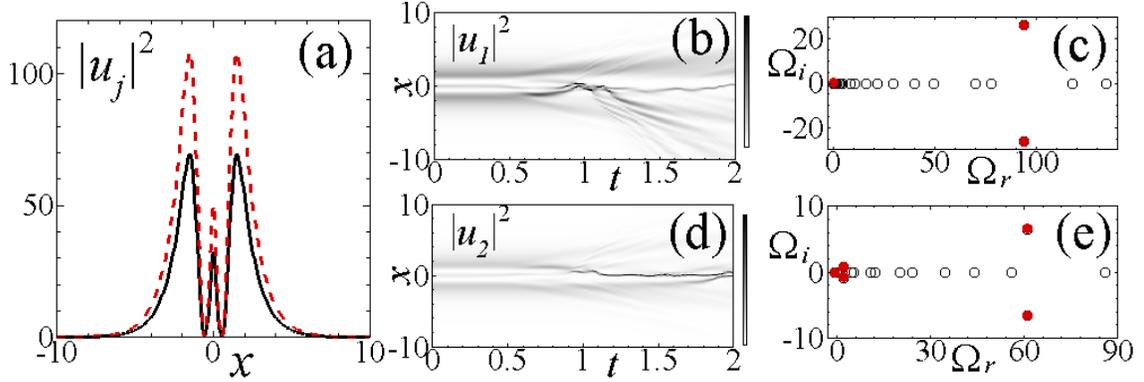,width=15 cm}
   \end{center}
\caption{Same as in Fig. \ref{dyn_BB1_3_-04} but for $n=3$ and  $g=0.4$.  
 \label{dyn_BB1_3_04}}
 \end{figure}

In the case of $n=3$  one more pare of complex conjugated eigenvalues appears in comparison with the case $n=2$ what corresponds to the more pronounced instability which starts to develop earlier then in previous case and in the case $g<0$ consists of transformation of the three-peaks profile of the solution into one-peak oscillating solution (see Figs.\ref{dyn_BB1_3_-04} and \ref{dyn_BB1_3_04}) while in the case $g>0$ the perturbed bright-bright solution disappears after strong radiation emission-

\section{Systems with quadratic  potentials}
\label{Sec4}

\subsection{Construction of families of solutions}

In this section, we construct explicit solutions of Eqs. (\ref{estacionario}) when either (or both) of the external potentials  $V_{1}$ or $V_{2}$ are different from zero.
We will consider the case when $\lambda_1=\lambda_2=\lambda$ and $E_1=E_2=0$.
By choosing $ c(x) = e^{ -\lambda x^{2}}$ and substituting into Eqs. (\ref{E1}) and  (\ref{E2}) we get quadratic trapping potentials of the form $V_1(x)=V_2(x) = \lambda^2 x^{2}$. 
These Gaussian nonlinearities can be generated by controlling the Feshbach resonances optically using a Gaussian beam  (see e.g. \cite{exper}). Here we also take $K=0$, thus
\begin{equation}\label{exa}
g_{jk}(x) = \tg_{jk} \exp\left( 3\lambda x^{2} \right),  \quad j,k=1,2.
\end{equation}
Our canonical transformation is given by 
$ X(x) =  \int_{0}^{x} ds \; \exp\left(\lambda s^{2}\right) = \frac12\sqrt{-\frac{\pi}{\lambda}}  \mathop{\mbox{erf}}\sqrt{-\lambda} x$. This leads us to a restriction on the sign of the chemical potential $\lambda$ which must be less then zero, $\lambda<0$.
 In this case Eq. (\ref{homogenea1}) becomes
\begin{subequations}\label{sistema2}
\begin{eqnarray}
  - \frac{d^{2}U_{1}}{dX^{2}} + \tg_{11}U_{1}^{3}+\tg_{12}U_{1}U_{2}^{2} &=& 0,\\
  - \frac{d^{2}U_{2}}{dX^{2}} + \tg_{22}U_{2}^{3}+\tg_{21}U_{1}^{2}U_{2} &=& 0.
\end{eqnarray}
\end{subequations}
Note that the range of $X$ is again finite since
$ -\tfrac12\sqrt{-\frac{\pi}{\lambda}} \le X \le \tfrac12\sqrt{-\frac{\pi}{\lambda}}$, 
and hence, we can again construct many localized solutions to Eq.  
(\ref{estacionario}) starting from solutions of Eqs. \eqref{sistema2} which satisfy the 
boundary conditions  $U_{j}( \pm\frac12\sqrt{-\frac{\pi}{\lambda}}) = 0$, $j=1,2$.

Using the condition  (\ref{cond}) we can reduce Eqs.  \eqref{sistema2} to
\begin{eqnarray}
\label{homogenea2_E0}
-\frac{d^{2}\Phi}{dX^{2}}+\sigma \Phi^{3}=0,
\end{eqnarray}
where we introduce new field $\Phi$ which is related with original fields through the relation (\ref{PhitoU})
and $\sigma={\rm sgn}[(\tg_{11}\tg_{22}-\tg_{12}\tg_{21})(\tg_{jj}-\tg_{12})]$.
For $\sigma=-1$ this equation has two solutions
\begin{subequations}
\label{sols}
\begin{eqnarray}
  \Phi^{(1)}(X) &=& \mu\cn(\mu X, k_{*}), \\
  \Phi^{(2)}(X) &=& \frac{\mu}{\sqrt 2}\frac{\sn(\mu X, k_{*})}{\dn(\mu X, k_{*})},
\end{eqnarray}
\end{subequations}
with $k_{*} = 1 / \sqrt{2}$, which solve the system  (\ref{sistema2}) and that $\Phi^{(1)}(X)$ and $\Phi^{(2)}(X)$  vanish when 
$\mu X = (2n+1) K\left( k_{*} \right)$ and
$\mu X = 2n K\left( k_{*} \right)$ correspondingly. 
Thus we come to an infinite number 
of solutions of system (\ref{sistema2}) under zero boundary conditions on the new finite interval, 
which correspond to different values of $\mu$.

Finally, localized solutions of the NLS system  (\ref{estacionario}), are given by 
\begin{eqnarray}
  u_{jn}(x) &=& 
   2n\sqrt{\frac{-\lambda}{\pi}} K\left( k_{*} \right) \theta_j \;
    e^{x^{2}/2 } \;
    \mathop{\mbox{cn}}\left( \chi_{n}(x), k_{*} \right),
  \quad n=1,3,...
\label{ex2-phi1}
\end{eqnarray}
and
\begin{eqnarray}
  u_{jn}(x) &=& 
    \sqrt{\frac{-2\lambda}{\pi}}   n K\left( k_{*} \right) \theta_j \;
    e^{x^{2}/2 } \;
     \frac{\textstyle \mathop{\mbox{sn}} \left( \chi_{n}(x), \, k_{*} \right) }
         {\textstyle \mathop{\mbox{dn}} \left( \chi_{n}(x), \, k_{*} \right) },
  \quad n=2,4,...
\label{ex2-phi2}
\end{eqnarray}
with
\begin{equation}
  \chi_{n}(x) = n K\left( k_{*} \right) \mathop{\mbox{erf}}\sqrt{-\lambda} x .
\end{equation}
It can be shown by simple asymptotic analysis that the last factors in Eqs. (\ref{ex2-phi1}) and (\ref{ex2-phi2}) tend to zero as $x \to \pm\infty$ faster than $\exp( -x^{2}/2 )$ and that 
these are indeed \emph{localized} solutions of our problem. 

\subsection{Dynamics and stability of the solutions}
 
\begin{figure}
\begin{center}
\epsfig{file=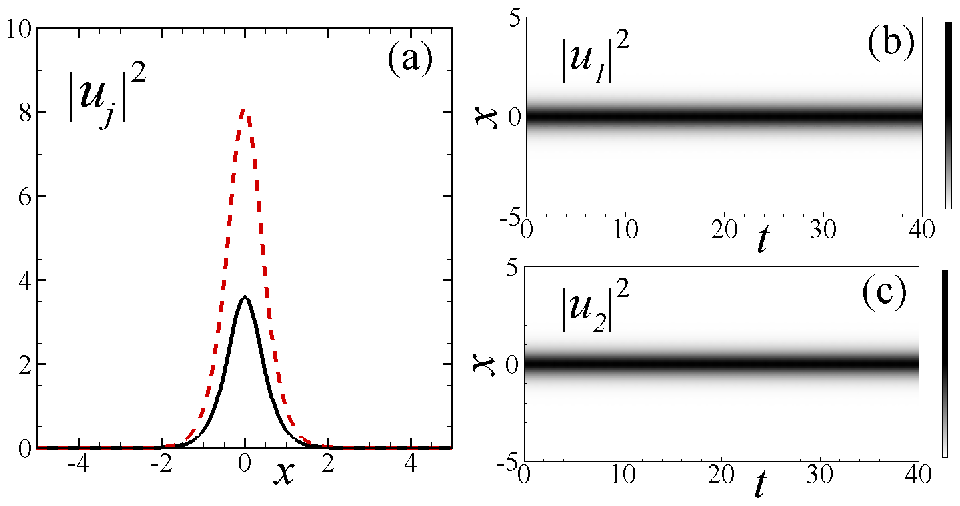,width=12 cm}
\end{center}
 \caption{Dynamics of a $n=1$ bright-bright soliton solution (\ref{ex2-phi1}) in a parabolic potential $V_j(x)=\lambda^2x^2$, ($j=1,2$) under a perturbation of the form (\ref{pert}). Shown are (a) the initial density profiles of the first (black solid) and second (red dashed) component. (b,c) Density plot of the time evolution of (b) $ |u_1(x,t)|^2$ (c) $ |u_2(x,t)|^2$.
Parameter values are $g=-0.1$, $g_{11}=-1$, $g_{22}=-0.5$, $\lambda=-1$, $\epsilon=0.001$, $q=0.2$.
 \label{dyn_BB1_V_-01}}
 \end{figure}

\begin{figure}
\begin{center}
\epsfig{file=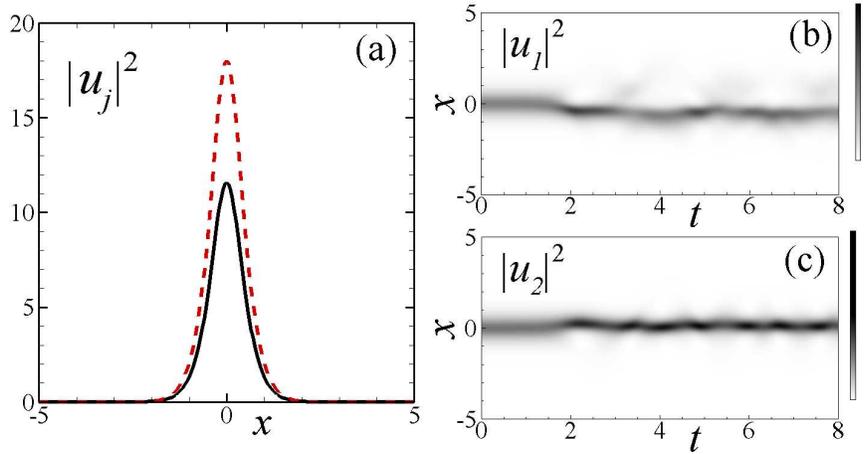,width=12 cm}
\end{center}
 \caption{Same as in Fig. \ref{dyn_BB1_V_-01} but for $g=0.4$.
\label{dyn_BB1_V_04}}
 \end{figure}

Let us now check the stability of the bright-bright soliton solutions 
%which can be constructed 
in the presence of the parabolic potential.
First we consider  solutions given by Eq. (\ref{ex2-phi1}) with $n=1$. 
As in the case of the bright-bright solution without external potential from Section~\ref{bb} here solution for $n=1$ is also linearly stable in all the domain of existence of the solution. 
However, by checking the dynamical stability under small perturbation of the initial exact solution we found that for fixed values of intra-species interactions $g_{11}=-1$ and $g_{22}=-0.5$ there is the domain of stability $\sqrt{g_{11}g_{22}}<g\leq 0$ (see an example in  Fig. \ref{dyn_BB1_V_-01}) and also that there is a domain where the solution is unstable $g>0$ (an example in  Fig. \ref{dyn_BB1_V_04}). As it is clear from  Fig. \ref{dyn_BB1_V_04}(b), (c) there is a phase separation effect related to the immiscibility of the two-component mixture. In this case the fist component is shifted to negative $x$ values while the second component is displaced to  positive $x$ values; however this separation rather small and components continue to have a significant overlapping.
 
\begin{figure}
\begin{center} 
\epsfig{file=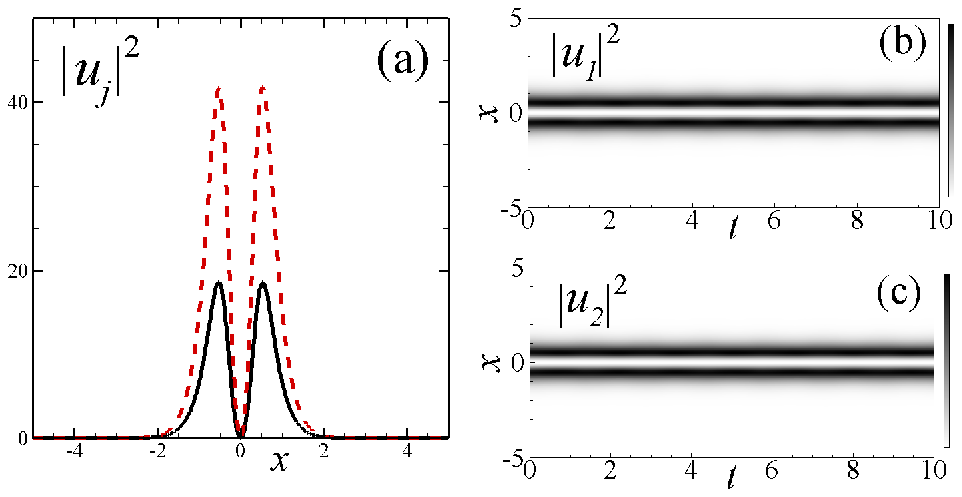,width=12 cm}
\end{center}
 \caption{Dynamics of a $n=2$ bright-bright soliton solution (\ref{ex2-phi2}) in a parabolic potential $V_j(x)=\lambda^2x^2$, ($j=1,2$) under a perturbation of the form (\ref{pert}). Shownare (a) the initial density profiles of the first (black solid) and second (red dashed) component. (b,c) Density plot of the time evolution of (b) $ |u_1(x,t)|^2$ (c) $ |u_2(x,t)|^2$.
Parameter values are $g=-0.1$, $g_{11}=-1$, $g_{22}=-0.5$, $\lambda=-1$, $\epsilon=0.001$, $q=0.2$.
\label{dyn_BB2_V_-01}}
 \end{figure}

\begin{figure}
\begin{center} 
\epsfig{file=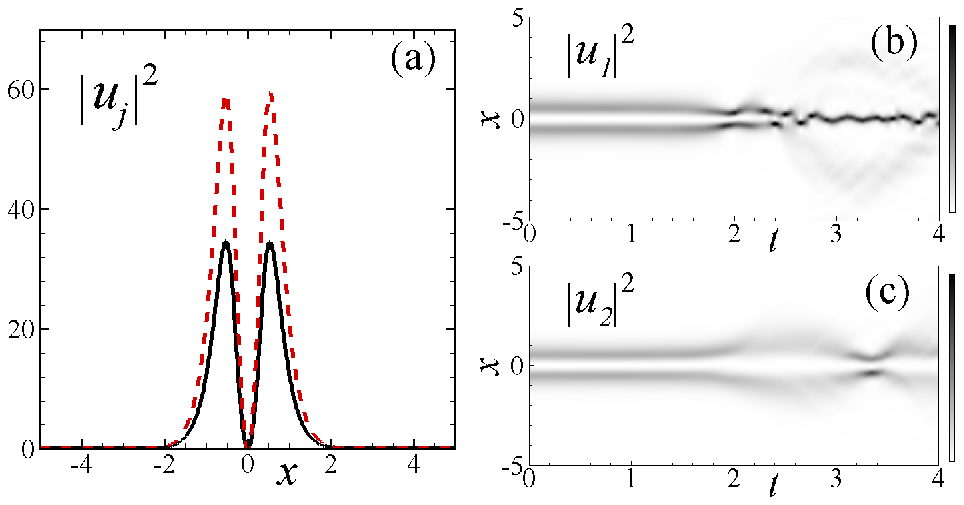, width=12 cm}
\end{center}
 \caption{Same as in Fig. \ref{dyn_BB1_V_-01} but for $g=0.2$.
 \label{dyn_BB2_V_02}}
 \end{figure}

Now let us consider stability  of the bright-bright solution for $n=2$. In contrast to the case without linear potential studied previously, here the solution is linearly stable in all of its domain of existence. From the dynamical stability analysis we find that, in the domain $\sqrt{g_{11}g_{22}}<g\leq 0$, the solution is dynamically stable while for $g>0$ as in the previous case of the $n=1$ solution  becomes unstable due to the immiscibility condition between the two components. Examples are shown in  Fig. \ref{dyn_BB2_V_-01} that has a stable dynamics and in Fig. \ref{dyn_BB2_V_02} where unstable behaviour of the bright-bright solution for $n=2$ is observed. In contrast to the case of bright-bright solution without external potential in the case at hand the second component due to reflection from the parabolic potential starts to oscillate.
 
\begin{figure}
\begin{center} 
\epsfig{file=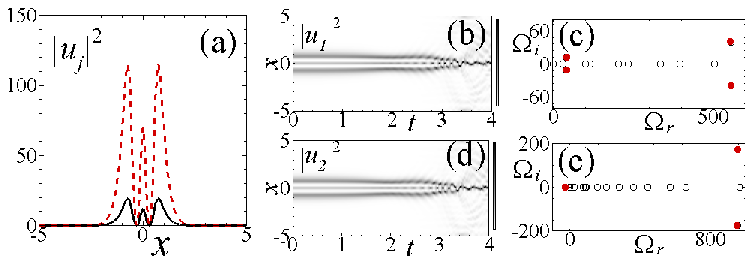,width=14 cm}
\end{center}
 \caption{Dynamics of a $n=3$ bright-bright soliton solution (\ref{ex2-phi1}) in a parabolic potential $V_j(x)=\lambda^2x^2$, ($j=1,2$) under a perturbation of the form (\ref{pert}). Shownare (a) the initial density profiles of the first (black solid) and second (red dashed) component. (b,c) Density plot of the time evolution of (b) $ |u_1(x,t)|^2$ (c) $ |u_2(x,t)|^2$.
Parameter values are $g=-0.4$, $g_{11}=-1$, $g_{22}=-0.5$, $\lambda=-1$, $\epsilon=0.001$, $q=0.2$.
 \label{dyn_BB3_V_-04}}
 \end{figure}

\begin{figure}
\begin{center} 
\epsfig{file=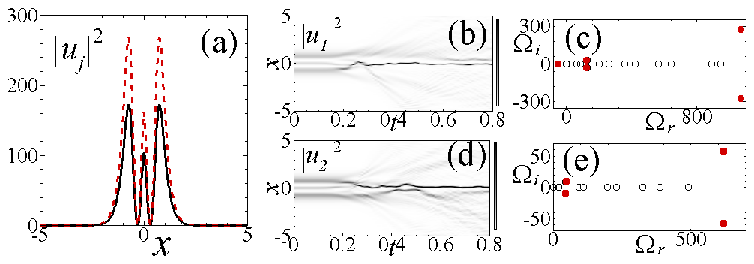,width=14 cm}
\end{center}
 \caption{Same as in Fig.\ref{dyn_BB3_V_-04} but for $g=0.4$.
 \label{dyn_BB3_V_04}}
 \end{figure}

Finally, when moving to solitons with $n=3$ we observe that in the full domain of existence of the bright-bright solution, it is linearly as well as dynamically unstable (see  Figs. \ref{dyn_BB3_V_-04} and \ref{dyn_BB3_V_04}).
 
\section{Dark-bright soliton solutions}
\label{Sec5}

\subsection{Dark-bright soliton solutions}

Up to now we have focused our attention in constructing vector extensions of the solutions known for the scalar case with spatially dependent nonlinear coefficients and studying their stability. In this section we will consider nontrivial solutions where the functional form of both components is completely different.

Dark-bright soliton solutions have been extensively studied in the context of BEC after the pioneering work of Busch and Anglin \cite{sol5}, where the coefficients $\tg_{ij}=1$, $j,k=1,2$, what corresponds to the well-known integrable Manakov system \cite{sol0}.
 
As in the dark-dark case to constract dark-bright solution it is netural to consider $V_{1}(x)=0$ for the dark component,  $\lambda_1=\lambda_2=\lambda=\omega^{2}/4>0$ and correspondingly $c(x)=1 + \alpha \cos \omega x$.
From Eq. (\ref{homogenea1}) we get the relation
\begin{subequations}
\begin{eqnarray}
\label{Energyrelations1}
E_1&=&\lambda(1-\alpha^2),\\
E_{2}&=&E_{1}-V_2(x)c(x)^2,%>0
\end{eqnarray}
\end{subequations}
and taking the potential $V_{2}(x)$ of the form
\begin{equation}\label{potencial2}
V_{2}(x)=K(1+\alpha\cos(\omega x))^{-2},
\end{equation}
from (\ref{Energyrelations1}) we get that $E_{1}\neq E_{2}$,  and
\begin{equation}K=E_{1}-E_{2}.
\end{equation}
Therefore, we can choose as solutions of Eqs.  (\ref{homogenea1}) the following dark-bright soliton solutions:
\begin{subequations}\label{DBDB}
\begin{eqnarray}
U_{1}(X)&=&\sqrt{\frac{E_{1}}{\tg_{11}}}\tanh(\mu X),\\
U_{2}(X)&=&\sqrt{\frac{E_1}{\tg_{11}}\frac{(\tg_{21}-\tg_{11})}{(\tg_{22}-\tg_{12})}}\sech(\mu X),
\end{eqnarray}
\end{subequations}
where 
\begin{subequations}
\begin{eqnarray}
\label{mu}
\mu&=&\sqrt{\frac{E_1}{2\tg_{11}}\frac{(\tg_{11}\tg_{22}-\tg_{12}\tg_{21})}{(\tg_{22}-\tg_{12})}},\\
\label{Energyrelations2}
 E_{2}&=&\frac{\tg_{21}}{\tg_{11}}E_{1}-\mu^{2}.
\end{eqnarray}
\end{subequations}
As it is clear in order to have real solutions one has to satisfy one of the following sets of conditions for the nonlinear coefficients: either $\tg_{22}>\tg_{12}$, $\tg_{21}>\tg_{11}$, $\tg_{11}\tg_{22}>\tg_{12}\tg_{21}$ or $\tg_{22}<\tg_{12}$, $\tg_{21}<\tg_{11}$, $\tg_{11}\tg_{22}<\tg_{12}\tg_{21}$. 

Then, the dark-bright type solution of Eq. (\ref{estacionario}) is given by
\begin{subequations}\label{dark_bright}
\begin{eqnarray}
u_{1}(x)&=&(1+\alpha\cos(\omega x))^{1/2} U_1(X(x)),\\
u_{2}(x)&=&(1+\alpha\cos(\omega x))^{1/2} U_2(X(x)),
\end{eqnarray}
\end{subequations}
where $X(x)$ is given by Eq. \eqref{cuci}. 

\subsection{Stability of dark-bright soliton solutions}

Dark-bright solutions given by Eqs. \eqref{dark_bright} are remarkably robust to perturbations as one would in principle expect. All of our simulations of perturbed dark-bright solitons have lead to stable behavior. 

\begin{figure}
\begin{center}
\epsfig{file=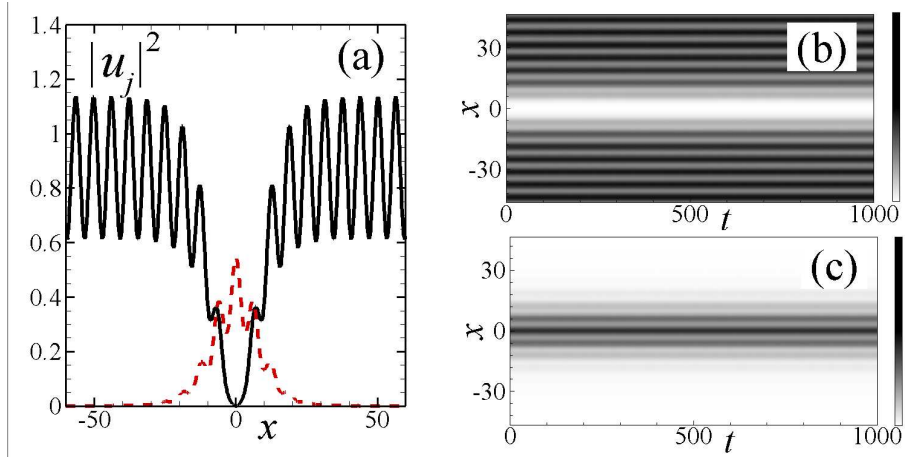,width=12 cm}
\end{center}
 \caption{Dynamics of a dark-bright soliton solution given by Eq. (\ref{dark_bright}) under a perturbation of the form (\ref{pert}). Shown are (a) the initial density profiles of the dark (black solid) and bright (red dashed) components, and ensity plots of the time evolution of (b) $ |u_1(x,t)|^2$ (c) $ |u_2(x,t)|^2$.
Parameter values are  $g_{11}=0.26$, $g_{22}=1$, $g=0.5$, $\alpha=0.3$, $\omega=1$, $\epsilon=0.02$, $q=0.2$. \label{dyn_DB_026}}
 \end{figure}

\begin{figure}
\begin{center}
\epsfig{file=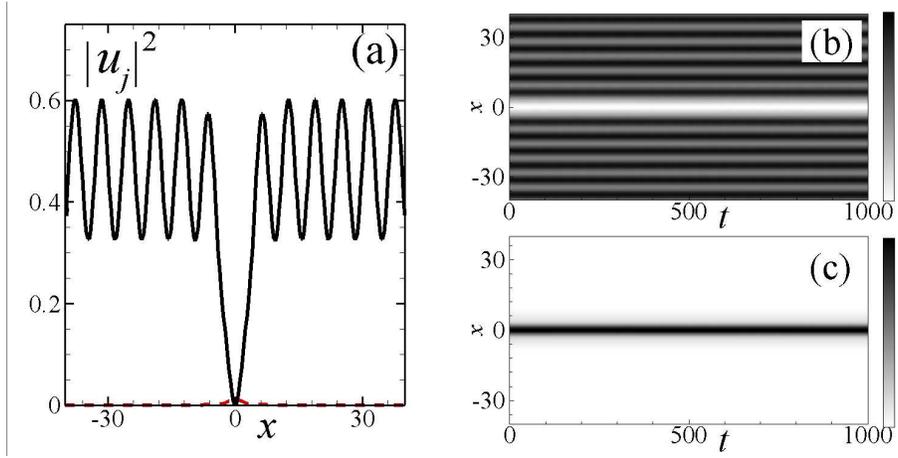,width=12 cm}
\end{center}
\caption{Same as in Fig. \ref{dyn_DB_026} but for $g_{11}=0.49$.
\label{dyn_DB_049}}
 \end{figure}

A couple of examples are shown in Figs. \ref{dyn_DB_026} and \ref{dyn_DB_049}  for two different values of the nonlinearity $g_{11}=0.26$ (Fig. \ref{dyn_DB_026}) and $g_{11}=0.49$ (Fig. \ref{dyn_DB_049}).

 \section{Conclusions}
\label{Sec6}

In conclusion, we have used Lie symmetries and canonical transformations to construct explicit solutions of  coupled nonlinear Schr\"odinger systems with spatially inhomogeneous nonlinearities starting from those of spatially homogeneous coupled nonlinear Schr\"odinger equations. 

Although we have restricted our attention to a few selected examples of physical relevance, the range of nonlinearities and potentials for which this can be done is very wide. We have constructed explicit solutions for $V=0$ with localized and periodic nonlinearities and discussed the parameter regimes for which solutions are stable. A similar methodology has been applied to construct and study the stability of specific families of solutions when the potential is quadratic on the space coordinates. Finally we have constructed dark-bright soliton solutions that are not direct extensions of known solutions of the scalar spatially inhomogeneous case.

\section*{Acknowledgements}

J. B.-B. would like to thank to P. J. Torres for discussions and for hospitality during his visit to the University of Granada.
This work has been partially supported by grants FIS2006-04190 
(Ministerio de Educaci\'on y Ciencia, Spain), PRINCET and PCI-08-0093 (Consejer\'{\i}a de Educaci\'on y Ciencia de la Junta de Comunidades de Castilla-La Mancha, Spain). 
V.A.B. acknowledges support from the FCT grant,
PTDC/FIS/64647/2006.

\appendix
\section{Linear stability analysis}
   \label{ApenA}
   
Let us  look for perturbed solutions of Eq. \eqref{GP}
\begin{eqnarray}
u_{j}(x,t)&=&\left( u_{0j} + A_j(x,t) \right)e^{i\lambda_j t},
\label{psi_pert}
\end{eqnarray}
where $u_{0j}$ are real solutions of the stationary problem (\ref{estacionario}) and $|A_{j}|\ll u_{0j}$ are perturbations. Substituting (\ref{psi_pert}) in  Eq. (\ref{estacionario}) and linearizing it with respect to $A_{j}(x)$ we obtain
\begin{multline}
i\dot A_{j}= \left(\lambda_j + V(x)\right) A_{j} - \nabla^2  A_{j}   +
\left(g_{jj}|u_{0j}|^2 +g |u_{0(3-j)}|^2\right)A_{j}    + \\
g_{jj} |u_{0j}|^2 \bar{A}_{j} +  2g u_{0j} u_{0(3-j)} \left(A_{3-j} + \bar{A}_{3-j}\right)    
 \label{a_lin} 
\end{multline}
Decomposing $A_{j}$ into its real and imaginary parts, $A_{j}(x,t) = P_{j}(x,t) + iQ_{j}(x,t)$, 
defining the operators
\begin{subequations}
\begin{eqnarray}
\label{operators}
L_j^{(+)}&=&\lambda_j-\nabla^2+3g_{jj}\left(u_{0j}\right)^2 + g\left(u_{0(3-j)}\right)^2, \\
L_j^{(-)}&=&\lambda_j-\nabla^2+g_{jj}\left(u_{0j}\right)^2 + g\left(u_{0(3-j)}\right)^2,
\end{eqnarray}
\end{subequations}
and introducing the notation $S(x)=2g u_{01}(x) u_{02}(x)$, we obtain
\begin{eqnarray}
\label{eigen_matrix}
\left( \begin{array}{c}
\dot P_{j}\\
-\dot Q_{j} \end{array} \right)&=&
\left( \begin{array}{cc}
0 & L_j^{(-)}\\
L_j^{(+)} & 0 \end{array} \right)
\left( \begin{array}{c}
P_{j}\\
Q_{j} \end{array} \right) 
+
\left( \begin{array}{cc}
0 & 0\\
S(x) & 0 \end{array} \right)
\left( \begin{array}{c}
P_{3-j}\\
Q_{3-j} \end{array} \right).
\end{eqnarray}

Looking for solutions in the form $P_{j}(x,t)=p_{j}(x) e^{w t}$ and $Q_{j}(x,t)=q_{j}(x) e^{w t}$, and defining $\Omega=-w^2$, Eqs. (\ref{eigen_matrix}) reduce to
\begin{eqnarray}
\label{eigen_matrix_simpl}
\Omega \left( \begin{array}{c}
p_{1}\\
p_{2} \end{array} \right) =
\left( \begin{array}{cc}
L_1^{(-)}L_1^{(+)} & L_1^{(-)}S(x)\\
L_2^{(-)}S(x) & L_2^{(-)}L_2^{(+)} \end{array} \right)
\left( \begin{array}{c}
p_{1}(x)\\
p_{2}(x) \end{array} \right).
\end{eqnarray}
Thus, a solitary wave solution $u_{0,j}$ will be linearly unstable when either $\Omega_r={\rm Re}(\Omega)< 0$ or $\Omega_i={\rm Im}(\Omega)\ne 0$.

  \end{document}